\documentclass[runningheads]{llncs}
\usepackage{graphicx}
\usepackage{amsmath}
\usepackage{bm}
\usepackage{subfigure}
\usepackage{color}
\usepackage{booktabs}

\usepackage{amssymb}

\usepackage{mathrsfs}
\usepackage{array}
\usepackage[marginal]{footmisc}

\begin{document}
\title{Sequential Tag Recommendation}

\author{
    Bing Liu\thanks{Equal Contribution.}\and
    Pengyu Xu\textsuperscript{*}\and
    Sijin Lu \and 
    Shijing Wang \and \\
    Hongjian Sun \and
    Liping Jing\thanks{Corresponding author.}
}

\authorrunning{Bing Liu et al.}

\institute{Beijing Key Lab of Traffic Data Analysis and Mining, Beijing Jiaotong University}% \institute{1 Beijing Key Lab of Traffic Data Analysis and Mining, Beijing Jiaotong University}

\maketitle              % typeset the header of the contribution

\begin{abstract}
With the development of Internet technology and the expansion of social networks, online platforms have become an important way for people to obtain information. The introduction of tags facilitates information categorization and retrieval. Meanwhile, the development of tag recommendation systems not only enables users to input tags more efficiently, but also improves the quality of tags. However, current tag recommendation methods only consider the content of the current post and do not take into account the influence of user preferences. Since the main body of tag recommendation is the user, it is very necessary to obtain the user's tagging habits. Therefore, this paper proposes a tag recommendation algorithm (MLP4STR) based on the dynamic preference of user's behavioral sequence, which models the user's historical post information and historical tag information to obtain the user's dynamic interest changes. A pure MLP structure across feature dimensions is used in sequence modeling to model the interaction between tag content and post content to fully extract the user's interests. Finally tag recommendation is performed.

\end{abstract}
\section{Introduciton}
With the development of Internet technology and the expansion of social networks, online platforms have become an important way for people to obtain information. The combination of digital media technology and Internet technology has given rise to a series of excellent knowledge service websites, such as Zhihu, Stack Exchange and Stack Overflow, which are question and answer websites. These Q\&A sites allow users to ask their own questions to seek relevant answers from other users, as well as to search for existing questions to gain knowledge. However, with the explosion of information on the Internet, it created a serious information overload problem for users. In order to solve this problem, the tagging mechanism was introduced to facilitate information categorization and retrieval~\cite{ref_article1}. Tag recommendation systems have emerged, which not only facilitate users to input tags but also improve the quality of tags~\cite{ref_article2}. In response to the needs of the industry, tag recommendation has become an important research topic that has attracted much attention in recent years, and a large amount of research work has already been done~\cite{ref_article3}. Existing tag recommendation algorithms can be roughly divided into two categories, namely content-based tag recommendation algorithms and collaborative filtering algorithms. The former directly classifies item text with multiple labels, while the latter discovers similarities between users based on their behaviors and interests, and then uses such similarities to predict items or content that users may like.

Tag recommendation involves information about three entities: users, items, and tags; if we ignore user information, then the tag recommendation problem can be modeled as a task of classifying items with multiple tags. Users can describe their opinions about items through tags, so tags are the link between users and items, and also an important data source to reflect users' interests. Moreover, labels are also keywords used by users to describe the focus of items, and due to the different living environments, hobbies, and research contents of users, they focus on the same issues differently.

% 加上图1
\begin{figure}
\includegraphics[width=\textwidth]{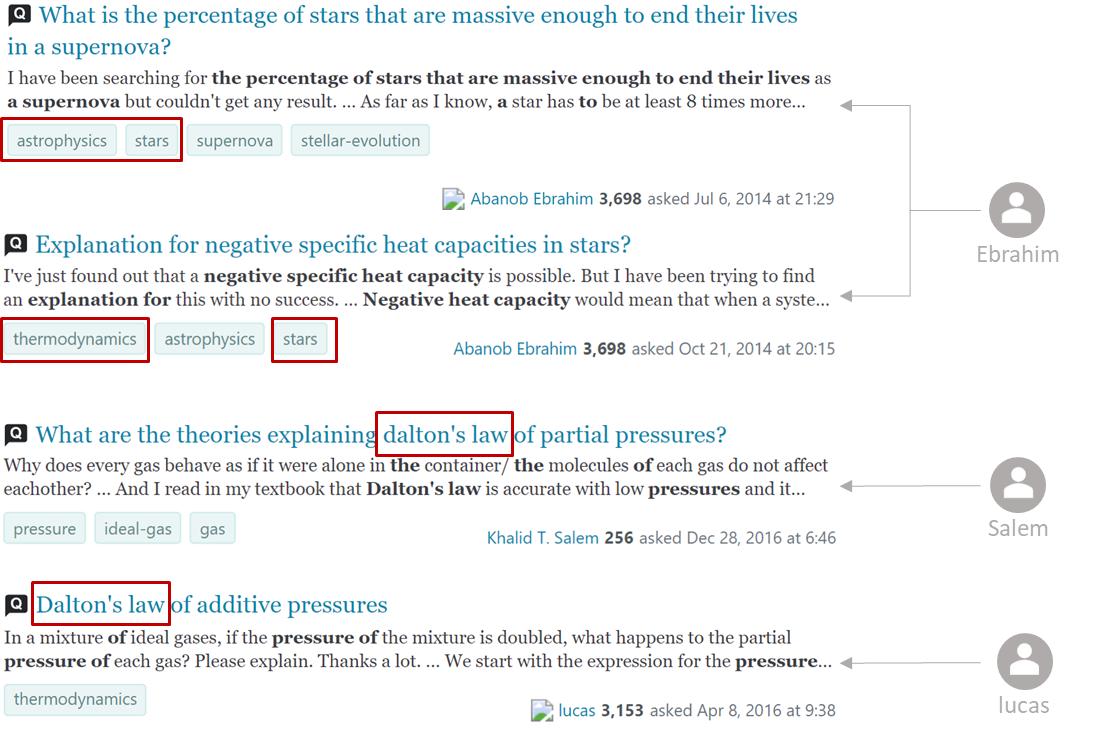}
\caption{Example of problems on Stack Exchange} \label{example QA}
\end{figure}

Fig.~\ref{example QA} shows four posts on Stack Exchange from the same user and different users, and it can be found that two posts published by the same user have overlapping parts of tags even though their contents are different, suggesting that the content of the user's research and the direction of his/her concern influence the user to add tags to the posts. In contrast, two identical posts about Dalton's law of partial pressure published by different users are labeled with completely different tags. This suggests that users' tagging choices are influenced by their personal subjective preferences at the time of posting. The ultimate goal of tag recommendation algorithms is to make the recommended tags acceptable and adoptable by users, so user preferences directly affect the accuracy of tag recommendations.

As an important influence factor, the introduction of user preference can play an extremely important role in tag recommendation. In addition, there is a sequential relationship between users' posts in the online platform, for example, in Figure 1, user Ebrahim posted two posts in June and September, which shows that the user is currently focusing on the contents related to "astrophysics" and "stars". As users' interests and research content are dynamic, the next preferences may differ from the current ones, but there is often a dependency relationship. Subsequent posts by Ebrahim in December were tagged with "gravity", "astrophysics", "earth", "thermal-radiation", "geophysics", where new tags appeared, but these tags were also correlated with the previous tags. Therefore, the dynamic change pattern of user interests is also reflected in the content of the posts posted by users and the labels they are labeled with. If this sequential dependency implicit in the content of the posts and the content of the tags can be captured, the user's recent preferences can be better captured for more accurate recommendations.

In this paper, we propose a tag recommendation algorithm based on dynamic preferences of user behavioral sequences. First, for subsequent sequence modeling, the text information is first feature extracted, and the pre-trained language model BERT is used to learn the textual features of the posts and obtain the semantic representations of the tags; then, the text and tag information are aligned, and the dynamic preferences of the users in the post content and tag content are captured through a pure MLP-based sequence modeling approach across feature dimensions, to obtain the post content and user preference representations that lie in the same space, and to fuse the preferences with the current post as feature representations. representation and user preference representation, and fusing this preference with the semantic features of the current post as a feature representation. Finally, cross entropy loss is utilized for training.

The main contributions of this paper are as follows:

\begin{itemize}
    \item In this paper, the sequential recommendation task is proposed for the first time. The traditional tag recommendation algorithm ignores the user subject position in the recommendation as well as the chronological order of posting, which is re-modeled as a sequential tag recommendation task in order to be more in line with the actual scenario. In addition, this paper collects and organizes datasets for related researches.
    \item For the above task, this paper proposes the algorithm MLP4STR: a sequence modeling approach using a cross-feature-aligned pure MLP that aligns the features of text and tags, based on which the learning of user sequences is performed.
    \item We perform experiments on four benchmark datasets. The F1@5 metric demonstrates an average improvement of 4.8\% over the best comparative algorithm, which further proves the effectiveness of our sequence modeling method.
\end{itemize}

\section{Related work}
The field of tag recommendation has now seen a significant amount of research work. Existing tag recommendation algorithms can be divided into two categories, namely content-based tag recommendation algorithms and collaborative filtering algorithms.

Content-based tag recommendation algorithms aim to directly model the content information of an item to recommend tags for the item, which tends to simplify the tag recommendation problem into a multi-label classification task, i.e., ignoring the user information, and recommending tags for the item by taking into account the information of the data related to the item's content as well as the correlation between the labels. It has been studied for various types of contents such as documents, images, songs, micro-videos and other contents. The focus of this paper is the case where the item content is text. Traditional content-based tag recommendation methods consider the textual content of items as bag of words (BoW) and design different tag ranking strategies for tag recommendation. For example, Song et al. constructed a two-part graph of documents and tags. After clustering, the tags are ranked according to their importance~\cite{ref_article4}. Xia et al. proposed a plain Bayesian-based multi-label classifier and used the predicted tagl ranking scores of the classifier ~\cite{ref_article5}.Wu et al. further considered the phenomenon of tag content relevance and used a probabilistic graph model to simulate the label generation process ~\cite{ref_article6}~\cite{ref_article7}. Due to the limitation of BoW characteristics, traditional methods are unable to obtain the sequential or spatial information of text content. Recently, researchers have employed deep learning architectures to capture this critical information. Gong et al. used a Convolutional Neural Network (CNN) with an attention mechanism consisting of global and local channels ~\cite{ref_article8}. Lei et al. further introduced a capsule network to encode the intrinsic spatial relationship between parts and the whole ~\cite{ref_article9}. On the other hand, Li et al. used a Long-Short term Memory (LSTM) model based on the topic attention mechanism, which integrates the topic distribution generated by the LDA topic model into sequence modeling ~\cite{ref_article10}. Considering the hierarchical structure of documents, ~\cite{ref_article12} uses Bi-directional Gated Recurrent Units (Bi-GRUs) and hierarchical attention mechanisms as sentence encoders. After the emergence of pre-trained language models, He et al. used pre-trained language models as encoders to encode text, which further improved the accuracy of tag recommendation algorithms [11].

Collaborative filtering algorithms include collaborative filtering of tags and collaborative filtering of users. Collaborative filtering of tags is to recommend tags by considering the correlation between tags. Because there exists a certain degree of correlation between the tags of the same item, for example, "astrophysics" and "stars" in the example in Fig.~\ref{example QA} have a strong correlation. Similar tag correlations are common in the tag sets corresponding to each item. Early traditional tag recommendation methods and deep learning-based tag recommendation methods have significantly improved the performance of tag recommendation by exploiting the correlations between tags. Collaborative filtering of users, on the other hand, takes the user's preferences into account for tag recommendation, i.e., personalized tag recommendation. There are some methods that can be used when performing tag recommendation. It is possible to directly connect the user's id information with the item content information and label information ~\cite{ref_article13}, and learn the potential user preferences through neural networks; however, at this time, the user information is very sparse and plays a very small role; after the introduction of the user's historical label information, it is possible to combine the id information with the label information, and construct a heterogeneous graph so as to obtain the representation of the user's preferences containing the label information ~\cite{ref_article14}; Alternatively, a self-encoder network can also be utilized to encode the historical tag information thereby obtaining the user preference representation ~\cite{ref_article15}; in addition to considering the user id information, the user historical tag information, the user historical item information can also be introduced into the mix, and an external storage unit storage is introduced in ~\cite{ref_article16} to store the user preferences learned by the attention network from the user's historical item information.

Accurate modeling of user behavior is an important research component of recommender systems. The goal of sequential recommender systems is to combine a personalized model of user behavior (based on historical activities) with the concept of "context" based on the user's recent behavior. Capturing useful patterns from sequential dynamics is challenging. Markov chains assuming that the next action is conditional only on the previous action (or several previous ones) are successfully used to characterize short-range item transitions for recommendation ~\cite{ref_article17}. Recurrent Neural Networks (RNNs) summarize all historical actions via hidden states used to predict the next action ~\cite{ref_article18}. Both of these methods, while performing well in specific situations, are only applicable to certain types of data. Markov chain-based methods, by making strong simplifying assumptions, perform well in highly sparse settings, but may not capture the complex dynamics of more complex scenes. Conversely, RNNs, while expressive, require large amounts of data (especially dense data) to perform well. Inspired by tansformer, a number of sequential recommendation models based on the self-attention mechanism ~\cite{ref_article19}~\cite{ref_article20}~\cite{ref_article21} have emerged to good effect. However, since self-attention and its ilk are insensitive to the order of the input items, they rely on additional processes, such as adding positional embeddings to the input sequences, in order to make the model aware of the information contained in the sequence order. However, existing self-attention methods combine item sequences and positional embeddings from two heterogeneous data types, potentially disrupting the underlying semantics of item embeddings ~\cite{ref_article22}. Second, the computational complexity of self-attentive methods is quadratic with the length of the input item sequence. Therefore, inspired by MLP-Mixer, Li et al ~\cite{ref_article23} introduced the structure of MLP-Mixer to propose a new sequential recommendation model based on MLP structure.

In order to make the tag recommendation algorithm better meet the user's needs, this paper introduces a variety of user's historical behavior information containing rich semantic information, and adopts the pure MLP structure across feature dimensions for sequence modeling to assist in tag recommendation.

\section{Proposed Method}
In this section the model proposed in this paper will be described in detail. Inspired by sequential recommendation, the user's preferences are obtained by sequence modeling the user's historical post sequence. The user preferences are then used to assist the tag recommendation of the current post. The model framework is shown in Fig.~\ref{model}:
% 图2 模型图
\begin{figure}
\includegraphics[width=\textwidth]{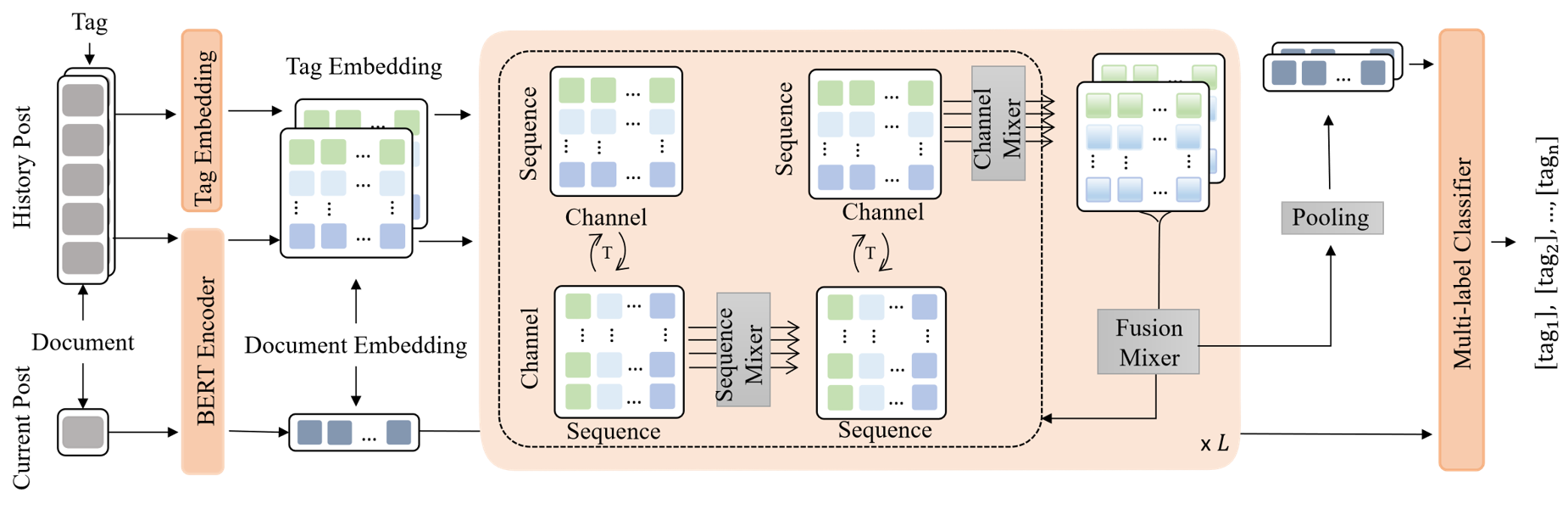}
\caption{Model framework} \label{model}
\end{figure}
\subsection{Problem Formulation}
In this paper, we consider the problem of tag recommendation when the project content is textual. We use fancy letters (e.g., $\mathcal{A}$) to denote sets, uppercase or lowercase letters (e.g., $A,a$) to denote scalars, lowercase bold letters (e.g., $\textbf{a}$) to denote vectors, and uppercase bold letters (e.g., $\textbf{A}$) to denote matrices. We denote the sample data input in the training phase as $P={\bm{(x_i,u_i,y_i)}}_{i=1}^N$, where $N$ denotes the number of sample posts, $x_i$ denotes the content of the $ith$ post, $\bm{u}_i$ denotes the author of the $ith$ post, and $\bm{y}_i=\{y_1,y_2,...,y_l,...,y_L \}$ denotes the set of candidate tags that may be involved in the $ith$ post. Here $y_l \in \{0,1\}$, if $y_l=1$ it means that the post is given the $lth$ label by the user; conversely, if $y_l=0$ it means that the post is not given the $lth$ label by the user. $L$ denotes the total number of tags in the candidate tag set. Each post can be represented as a sequence of words i.e. $\bm{x_i=(w_1,w_2,...,w_j,...,w_n)}$, where $\bm{w_j}$ denotes the one-hot encoded vector of the $jth$ word and n denotes the length of the word sequence of the post. The historical posts published by author $u_i$ are the sequence $\{\bm{D_1,D_2,...D_{u_i}}\}$, where $u_i$ is the number of posts published by $\bm{u_i}$. Historical posts include two parts of information, post content and tags, and the sequence of historical posts can be split into a text sequence $\bm{d}_i=\{d_1^i,d_2^i,... ,d_k^i,... ,d_{u_i}^i\}$ and the tag sequence $\bm{t}_i=\{t_1^i,t_2^i,,... ,t_k^i,... ,t_{u_i}^i\}$, where $d_k^i$ and $t_k^i$ denote the text content and the corresponding tag content of the $kth$ post published by user $\bm{u}_i$, respectively. In the testing phase we expect to be able to recommend tags related to the posts that meet the user's expectations for the posts that the user inputs in.

\subsection{Document Representation Learning}
Tag recommendation involves three entities: user, document and tag, and tag recommendation needs to learn a good document representation first. Pre-trained language modelin has made significant progress on natural language processing tasks. It is first pre-trained on massive text under unsupervised objectives and then fine-tuned on task-specific data.BERT has achieved good results in several natural language processing tasks, so here we apply BERT to this paper to learn the vector representation of the text, and then adapt it to the multi-labeled text categorization task model by fine-tuning. First construct the BERT input sequence $\bm{x_i}$ for the $ith$ post.
\begin{equation}
\bm{x}_i=\{[CLS],w_{i1},w_{i2},…,w_{i(n-2)},[SEP]\}
\end{equation}
where $w_{ij}$ denotes the jth word in the ith post, and [CLS] and [SEP] are the tag symbols located at the beginning and the end of the sequence, respectively. For convenience, we denote the length of the sequence as n. The text encoder encodes the input and represents it as:
\begin{align}
\bm{H_i}=\rm{BERT}(\bm{x_i}) \\
\bm{h_i} = \rm{avg\_pooling}(\bm{H}_i)
\end{align}
where $\bm{H}_i\in \mathbb{R}^{n\times d_h}$, $\bm{h_i} \in \mathbb{R}^{d_h}$ and $d_h$ is the word representation dimension.

\subsection{Tag Representation Learning}
In addition to historical posting information, historical tagging information is also important information about the user's historical behavior, so it is necessary to learn the representation of tagging content. The user's historical posting sequence involves a historical tagging sequence $\bm{t_i}=\{t_1,t_2,,... ,t_k,... ,t_{u_i}\}$, the implicit representation of labels needs to be learned before modeling the user history sequence $\bm{z}\in R^{L\times d_h}$. We obtain the representation of a label by aggregating the text encoding corresponding to the positive samples of the label:

\begin{equation}
\bm{z}_l=\frac{\bm{v}_l}{||\bm{v}_l||}, \bm{v_l}=\sum_{i:yi,l=1} \bm{h_i},l=1,…,L
\end{equation}

where $\bm{z}_l\in\bm{R}^{d_h}$, the embedding dimension of the tag is the same as the embedding dimension of the text. The document representation averaging process maintains the representation of the document and the representation of the tag in the same space. The tag representations obtained by the document representation averaging process represent the semantics of the tags well and implicitly take into account the correlation between the tags. The obtained tag representations will be used for subsequent user preference learning.

\subsection{User preference learning}
For user preference modeling, sequence modeling is used to model the historical behavior of users. To make full use of the user's id information, historical item information, and historical tagging information, the user's posting sequences and tagging sequences are modeled as sequences, and the sequence relationships among sequences, channels, and features are modeled by MLP-Mixer.

The user's historical postings contain rich preference information such as interests and hobbies. Making full use of the historical posting information can mine the hidden interests of users and the change of interest trends, which can be used as antecedent information to help understand the content of the posts published in the new phase. The tags corresponding to the user's historical posts are the user's summary of the content of the posted posts, which on the one hand is a condensed representation of the content of the historical posts, and on the other hand the user's historical used tags reflect the user's habit of tagging, especially corresponding to the content of the posts, which can reflect the user's habit of thinking. Therefore, the information in both parts is helpful for tagging recommendation of new posts. The approach in this paper explicitly models these two parts of information separately to capture user preferences, and uses a fusion layer to fuse the information of tags and posts to model the interaction between posts and the corresponding tagged content.

\subsubsection{Framework Introduction}

\begin{figure}
\centering
\includegraphics[width=0.6\textwidth]{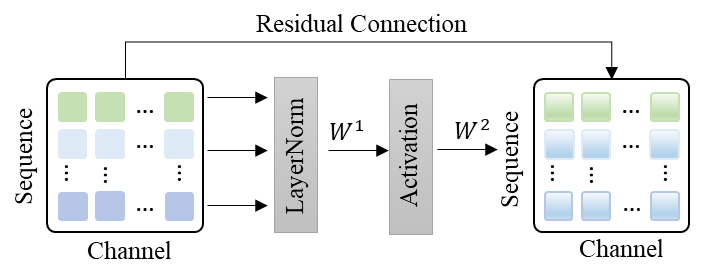}
\caption{Structure of the Sequence-mixer} \label{mixer}
\end{figure}

In this paper, a pure MLP-based structure is used to model sequence information. As shown in Fig.~\ref{mixer}, the historical posting sequences and historical tagging sequences are used as inputs to obtain the user's integrated preferences through integrated hybrid layers. Each integrated hybrid layer contains three parts. First is a sequence-mixer, which models sequence dependencies, i.e., the sequential relationships that exist between sequences composed of the same channels of each post or tag embedding in $D_i$; then a channel-mixer, which models the relationships that exist between different feature dimensions of the same post or tag embedding; and finally a fusion-mixer, which models the relationships that exist between different feature dimensions of the same post or tag embedding; and finally a fusion-mixer. mixer, which models the interaction between information from posts and information from tags.

\subsubsection{Structure of the Mixer}
% 这里换段出现了问题
\textbf{Sequence-mixer}: It is a MLP-based structure that learns sequential relationships in sequences of historical tags and sequences of historical text. Take the historical text sequence $\bm{d}_i=\{d_1^i,d_2^i, \dots ,d_k^i,\dots ,d_{u_i}^i\}$ for example, first padding the sequence according to the length u to get the sequence $\{d_1^i,d_2^i,... ,d_k^i,... ,d_u^i\}$, and then its encoded embedding represents the sequence $\bm{H}_d=\{\bm{h}_{d_1},\bm{h}_{d_2},... ,\bm{h}_{d_k},... ,\bm{h}_{d_u}\}$ as input, as shown in the table below, the dimensions of its embedded representation can all be represented as a sequence, for example, the $jth$ dimension can be represented as the sequence $\{\bm{h}_{d_1}^j,\bm{h}_{d_2}^j,... ,\bm{h}_{d_k}^j,... ,\bm{h}_{d_u}^j\}$, there is a temporal order relationship in this sequence, which can show the change of user's interest over time, and the order relationship is captured by the following structure:

\begin{equation}
{\hat{\bm{h}}}_{d}^t=\bm{h}_{d}^t+\bm{W}^2g(\bm{W}^1\rm{LayerNorm}(\bm{h}_{d}^t))
\end{equation}
where $t=1,2,\ldots,d_h,{\hat{\bm{h}}}_{d}^t,\bm{h}_{d}^t \in \mathbb{R}^u, \bm{W^1} \in \mathbb{R}^{r_u \times u} $denotes the weight matrix of the first fully connected layer, $\bm{W^2} \in  \mathbb{R}^{u \times r_u}$ denotes the weight matrix of the second fully connected layer, $r_u$ is the dimensionality of the adjustable hidden layer, and g denotes the nonlinear activation function.

% 图3：Sequence-mixer的结构

\textbf{Channel-mixer}: Like sequence-mixer, channel-mixer is a pure MLP block with a similar architecture; the key difference between them is their purpose. The goal of channel-mixer is to learn correlations within embedding vectors. Different information is expressed in different dimensions of textual features and labeled features, and after learning the sequence information, the correlation between the different feature dimensions is further learned
\begin{equation}
{\hat{\bm{h}}}_{d}^c=\bm{h}_{d}^c+\bm{W}^2g(\bm{W}^1\rm{LayerNorm}(\bm{h}_{d}^c))
\end{equation}
where $c=1,2,\ldots,u, {\hat{\bm{h}}}_{dc}, \bm{h}_{dc} \in \mathbb{R}^{d_h}, \bm{W^1} \in \mathbb {R}^{r_{d_h} \times d_h}$ denotes the weight matrix of the first fully-connected layer, $\bm{W^2} \in \mathbb{R}^{r_{d_h} \times d_h}$ denotes the weight matrix of the second fully-connected layer, $r_{d_h}$ is the dimensionality of the adjustable hidden layer, and g denotes the nonlinear activation function.

\textbf{Fusion-mixer}: The fusion mixer is a key component that connects text features with tag feature correspondences. Tags corresponding to different texts may have different meanings, and text contents with different labels may have different focuses, so fusion of the two parts can be used to get a more accurate embedding representation of the tag and text correspondences. Adopting the same structure as sequence mixer and channel mixer, a fusion is performed after each sequence mixing and channel mixing, and the fusion layer can be represented as:
\begin{equation}
{\hat{\bm{h}}}_{t_c}=\bm{h}_{t_c}+\bm{W}^2g(\bm{W}^1 \rm{LayerNorm}(\bm{h}_{t_c}))
\end{equation}
where $t=1,2,\ldots,d_h,c=1,2,\ldots,u, {\hat{\bm{h}}}_{t_c},\ \bm{h}_{t_c}\in\mathbb{R}^2$

\subsection{Tag Prediction}
Combining the document representation and user preference representation we obtained, the final tag prediction is performed. The probability of tag k is obtained by fusing different information through a classifier:
\begin{equation}
f_s(\bm{x}_i,\bm{u}_i, k)=\rm{sigmoid}(\alpha\bm{W}_0\bm{h}+\beta\bm{W}_1\rm{pooling}({\hat{\bm{H}}}_d^t)+\ \gamma\bm{W}_2\rm{pooling}({\hat{\bm{H}}}_d^c))
\end{equation}

Get the b tags with the highest probability among them as the recommended tags for the post:
\begin{equation}
f_b(\bm{x}_i,\bm{u}_i)=\rm{Top_b}([f_s(\bm{x}_i,\bm{u}_i,0),\ldots,,f_s(\bm{x}_i,\bm{u}_i,l)])
\end{equation}
where $f_b(\bm{x}_i,\bm{u}_i)$ is the set of b tags with the highest scores, i.e., the set of tags recommended for the post.

\subsection{Loss Function}
We use the corresponding binary cross-entropy function as a loss function for tag ranking to train the model parameters.
\begin{equation}
L=-\sum_{i=1}^{N}\sum_{k=1}^{l}{(\bm{y_{i,k}}\rm{log}{(f_s(\bm{x}_i,\bm{u}_i,k))})+(1-\bm{y_{i,k}})\rm{log}(1-f_s(\bm{x}_i,\bm{u}_i,k))}
\end{equation}
where N denotes the number of data in the training set, l denotes the number of labels, $\bm{y_{i,k}}\in\{0,1\}$ denotes whether the $ith$ post is assigned label k, and $\bm{y_{i,k}}=1$ denotes that the $ith$ post has label k, and vice versa.

\section{Experiments}
This section first describes the datasets and the baseline algorithms used for comparison, followed by performance evaluation and model validity analysis, respectively.
\subsection{Datasets}
In this paper, we use four real-world datasets: Physics, Academic, Cooking, and Android. These datasets are all from the StackExchange officially released and publicly available dataset website. Since user data with insufficient historical information does not fit the application object of our method, we filtered the user data according to the number of historical postings to get the final dataset used. The statistical information of the four datasets is shown in Table.~\ref{dataset}. When dividing the training set validation set and test set, in this paper, we use the last interaction post of the user interaction sequence as the test set, the penultimate interaction post as the validation set, and all the previous interacted posts as the training set ~\cite{ref_article27}.
% 表1 实验数据集
\begin{table}[htbp]
  \centering
  \caption{Summary of Experimental Datasets}
  \begin{tabular}{>{\centering\arraybackslash}p{5.5em} *{7}{>{\centering\arraybackslash}p{3.5em}}}
    \toprule
    \textbf{Dataset} & \multicolumn{1}{c}{\textbf{N}} & \multicolumn{1}{c}{\textbf{T}} & \multicolumn{1}{c}{\textbf{U}} & \multicolumn{1}{c}{$\hat{N}$} & \multicolumn{1}{c}{\textbf{W}} & \multicolumn{1}{c}{$\hat{W}$} & \multicolumn{1}{c}{$\hat{T}$} \\
    \midrule
    \textbf{physics} & 55716 & 883   & 4038  & 13.8  & 183042 & 206.6 & 3.2 \\
    \textbf{Academia} & 14062 & 439   & 1850  & 7.6   & 43247 & 188.1 & 2.7 \\
    \textbf{Cooking} & 10941 & 813   & 1338  & 8.2   & 35332 & 125.7 & 2.4 \\
    \textbf{Android} & 49724 & 1277  & 1911  & 26    & 49724 & 134.5 & 2.5 \\
    \bottomrule
  \end{tabular}
  \label{dataset}
  \\
  N is the number of posts, T is the number of tags, U is the number of users, $\hat{N}$ is the average number of posts by users, W is the total number of words, $\hat{W}$ is the average number of words, and $\hat{T}$ is the average number of tags
\end{table}

\subsection{Evaluation indicators}
We choose evaluation metrics that are widely used in the tag recommendation problem: Precision@k (P@k), Recall@k (R@k) and F1-score@k, defined as follows:
\begin{align}
P@K=\rm{avg}\frac{|{\hat{T}}_{u,i}\cap T_{u,i}|}{K}\\
R@K=\rm{avg}\frac{|{\hat{T}}_{u,i}\cap T_{u,i}|}{T_{u,i}} \\
F1@K=\frac{2\bullet P@K\bullet R@K}{P@K+R@K}
\end{align}
where $T_{u,i}$ denotes the label predicted for the ith post and ${\hat{T}}_{u,i}$ denotes the actual label corresponding to the ith post.

\subsection{Baselines}
To demonstrate the effectiveness of MLP4STR on the four datasets, we selected six most representative baseline models.

TLSTM ~\cite{ref_article10}: uses the LDA topic model to obtain topic information to enrich the representation of the document, thus using the topics to extract information from the document.

ABC ~\cite{ref_article24}: uses CNN neural network and word level attention mechanism to extract global and local information for tag recommendation.

CAN ~\cite{ref_article9}: incorporates an attention mechanism in capsule networks to capture important and distinguishable features.

PBAM ~\cite{ref_article25}: built a location-based attention model to automatically tag documents with keywords.

HAN ~\cite{ref_article26}: used a deep recurrent neural network to encode scientific and technical paper titles and abstracts into semantic vectors. The deep recurrent neural network in this uses Bidirectional Gated Recurrent Units (Bi-GRUs) with a hierarchical attention mechanism.

PTM4Tag ~\cite{ref_article11}: feature extraction using a pre-trained model and feature fusion of average pooled vectors.

\subsection{Parameter settings}
In this paper, we use Natural Language Toolkit (NLTK) to perform the word segmentation operation and keep the most frequently occurring 50,000 words as the vocabulary list. The hidden layer size of the model is set to 512 and the discard rate is 0.1. For all comparison algorithms, the word embedding dimension is set to 100, and the hidden layer size of GRU/LSTM is set to 256. In this paper, the Adam optimizer is used and the training process is stopped early when the validation loss is no longer reduced.

\subsection{performance comparison}
The evaluation metrics scores of the proposed MLP4STR and the six compared algorithms on the four datasets are shown in Table.~\ref{R_P},~\ref{R_Ac},~\ref{R_C} and~\ref{R_An} with the optimal results in bold and the sub-optimal results underlined.
% 表2-表5
\begin{table}[htbp]
  \centering
  \caption{Comparative results of evaluation metrics on the Physics dataset}
    \begin{tabular}{p{5.5em}|ccc|ccc|ccc}
    \toprule
    Method    & \multicolumn{1}{p{3.25em}}{P@1} & \multicolumn{1}{p{3.25em}}{R@1} & \multicolumn{1}{p{3.25em}|}{F1@1} & \multicolumn{1}{p{3.25em}}{P@3} & \multicolumn{1}{p{3.25em}}{R@3} & \multicolumn{1}{p{3.25em}|}{F1@3} & \multicolumn{1}{p{3.25em}}{P@5} & \multicolumn{1}{p{3.25em}}{R@5} & \multicolumn{1}{p{3.25em}}{F1@5} \\
    \midrule
    \textbf{TLSTM} & 52.3  & 20.2  & 29.2  & 36.6  & 39.7  & 38.1  & 27.9  & 48.9  & 35.5 \\
    \textbf{PBAM} & 60.2  & 23.2  & 33.5  & 41.6  & 44.9  & 43.2  & 31.2  & 54.9  & 39.8 \\
    \textbf{HAN} & 57.4  & 22.2  & 32    & 39.3  & 42.6  & 40.9  & 29.9  & 52.6  & 38.1 \\
    \textbf{ACN} & 54.3  & 20.7  & 30    & 36.1  & 39    & 37.5  & 27.3  & 48.2  & 34.9 \\
    \textbf{ABC} & 64.1  & 24.8  & 35.8  & 44.6  & 48.3  & 46.4  & 33.6  & 58.7  & 42.7 \\
    \textbf{PTM4Tag} & \underline{70.7}  & \underline{27.8}  & \underline{39.9}  & \underline{49.5}  & \underline{53.8}  & \underline{51.5}  & \underline{37.1}  & \underline{65}    & \underline{47.3} \\
    \textbf{MLP4STR} & \textbf{73.8} & \textbf{29.2} & \textbf{41.8} & \textbf{51.7} & \textbf{55.9} & \textbf{53.8} & \textbf{38.6} & \textbf{67.3} & \textbf{49} \\
    \bottomrule
    \end{tabular}%
  \label{R_P}%
\end{table}%

\begin{table}[htbp]
  \centering
  \caption{Comparative results of evaluation metrics on the Academic dataset}
    \begin{tabular}{p{5.5em}|ccc|ccc|ccc}
    \toprule
    Method   & \multicolumn{1}{p{3.25em}}{P@1} & \multicolumn{1}{p{3.25em}}{R@1} & \multicolumn{1}{p{3.25em}|}{F1@1} & \multicolumn{1}{p{3.25em}}{P@3} & \multicolumn{1}{p{3.25em}}{R@3} & \multicolumn{1}{p{3.25em}|}{F1@3} & \multicolumn{1}{p{3.25em}}{P@5} & \multicolumn{1}{p{3.25em}}{R@5} & \multicolumn{1}{p{3.25em}}{F1@5} \\
    \midrule
    \textbf{TLSTM} & 43.8  & 17.9  & 25.4  & 30.3  & 36.4  & 33    & 22.9  & 45.2  & 30.4 \\
    \textbf{PBAM} & 48.3  & 19.8  & 28    & 31.8  & 37.4  & 34.4  & 24.4  & 47.3  & 32.2 \\
    \textbf{HAN} & 42.2  & 17.1  & 24.3  & 27.6  & 32.4  & 29.8  & 21.4  & 41.5  & 28.2 \\
    \textbf{ACN} & 38.1  & 15.1  & 21.6  & 26    & 30.1  & 27.9  & 20.1  & 38.5  & 26.4 \\
    \textbf{ABC} & 58.2  & 24.2  & 34.2  & 39.8  & 47.5  & 43.3  & 30.2  & 58.7  & 39.8 \\
    \textbf{PTM4Tag} & \underline{62.2}  & \underline{26}    & \underline{36.7}  & \underline{42.6}  & \underline{50.2}  & \underline{46.1}  & \underline{31.5}  & \underline{60.9}  & \underline{41.5} \\
    \textbf{MLP4STR} & \textbf{65.8} & \textbf{27.6} & \textbf{38.8} & \textbf{44.4} & \textbf{52.8} & \textbf{48.2} & \textbf{32.9} & \textbf{64} & \textbf{43.4} \\
    \bottomrule
    \end{tabular}%
  \label{R_Ac}%
\end{table}%

\begin{table}[htbp]
  \centering
  \caption{Comparative results of evaluation metrics on the Cooking dataset}
    \begin{tabular}{p{5.5em}|ccc|ccc|ccc}
    \toprule
    Method    & \multicolumn{1}{p{3.25em}}{P@1} & \multicolumn{1}{p{3.25em}}{R@1} & \multicolumn{1}{p{3.25em}|}{F1@1} & \multicolumn{1}{p{3.25em}}{P@3} & \multicolumn{1}{p{3.25em}}{R@3} & \multicolumn{1}{p{3.25em}|}{F1@3} & \multicolumn{1}{p{3.25em}}{P@5} & \multicolumn{1}{p{3.25em}}{R@5} & \multicolumn{1}{p{3.25em}}{F1@5} \\
    \midrule
    \textbf{TLSTM} & 31.5  & 15    & 20.3  & 19.7  & 27    & 22.8  & 14.9  & 33.4  & 20.6 \\
    \textbf{PBAM} & 32.4  & 15.3  & 20.8  & 20.4  & 27.8  & 23.5  & 15.4  & 34.2  & 21.2 \\
    \textbf{HAN} & 20    & 8.9   & 12.4  & 13.7  & 18.3  & 15.7  & 10.7  & 23.6  & 14.7 \\
    \textbf{ACN} & 19.3  & 8.7   & 12    & 12.3  & 16.3  & 14    & 9.5   & 20.9  & 13.1 \\
    \textbf{ABC} & 60.5  & 29    & 39.2  & 37    & 50.5  & 42.7  & 26.3  & 58.4  & 36.2 \\
    \textbf{PTM4Tag} & \underline{69.1}  & \underline{33.7}  & \underline{45.3}  & \underline{41.9}  & \underline{57.1}  & \underline{48.3}  & \underline{28.8}  & \underline{63.9}  & \underline{39.7} \\
    \textbf{MLP4STR} & \textbf{72.6} & \textbf{35.5} & \textbf{47.7} & \textbf{44} & \textbf{60} & \textbf{50.8} & \textbf{30.9} & \textbf{68.4} & \textbf{42.6} \\
    \bottomrule
    \end{tabular}%
  \label{R_C}%
\end{table}%

\begin{table}[htbp]
  \centering
  \caption{Comparative results of evaluation metrics on the Android dataset}
    \begin{tabular}{p{5.5em}|ccc|ccc|ccc}
    \toprule
    Method    & \multicolumn{1}{p{3.25em}}{P@1} & \multicolumn{1}{p{3.25em}}{R@1} & \multicolumn{1}{p{3.25em}|}{F1@1} & \multicolumn{1}{p{3.25em}}{P@3} & \multicolumn{1}{p{3.25em}}{R@3} & \multicolumn{1}{p{3.25em}|}{F1@3} & \multicolumn{1}{p{3.25em}}{P@5} & \multicolumn{1}{p{3.25em}}{R@5} & \multicolumn{1}{p{3.25em}}{F1@5} \\
    \midrule
    \textbf{TLSTM} & 15.2  & 7.1   & 9.7   & 9.9   & 13.7  & 11.5  & 7.7   & 17.4  & 10.7 \\
    \textbf{PBAM} & 21.4  & 10    & 13.6  & 14    & 19    & 16.1  & 10.8  & 23.7  & 14.8 \\
    \textbf{HAN} & 14    & 6.4   & 8.8   & 10.5  & 14.1  & 12.1  & 8.3   & 18.1  & 11.3 \\
    \textbf{ACN} & 15.1  & 7     & 9.6   & 10.4  & 13.6  & 11.8  & 8.3   & 17.8  & 11.3 \\
    \textbf{ABC} & 45.1  & 21.8  & 29.4  & 27.3  & 37.2  & 31.5  & 19.8  & 43.9  & 27.3 \\
    \textbf{PTM4Tag} & \underline{54.9}  & \underline{26.5}  & \underline{35.7}  & \underline{32.3}  & \underline{43.3}  & \underline{37}    & \underline{22.9}  & \underline{49.7}  & \underline{31.3} \\
    \textbf{MLP4STR} & \textbf{56.3} & \textbf{27.4} & \textbf{36.8} & \textbf{33.9} & \textbf{45.7} & \textbf{38.9} & \textbf{23.7} & \textbf{51.7} & \textbf{32.5} \\
    \bottomrule
    \end{tabular}%
  \label{R_An}%
\end{table}%

From the comparison of experimental results, it can be seen that MLP4STR is significantly better than the existing methods. For example, the improvement over the best competing method on F1@5 is 3.6\%, 4.6\%, 7.3\%, and 3.8\% on four datasets, respectively. TLSTM introduces static latent topics for tag recommendation, but topic generation and tag recommendation are separated, so these topics are not necessarily suitable for the tag recommendation task. ABC, PBAM, and ACN are all attention-based neural networks that utilize local and global attention information for tag recommendation. PTM4Tag uses pre-trained models to learn the features of different modalities, and the features of different modalities are first averaged pooled and then fused. Although the above methods design different text representation strategies, they only use the features corresponding to the current post, and do not mine the preferences displayed by the user's historical posts. Firstly, MLP4STR uses a pre-trained language model to learn the representation of text features, and then obtains the user's historical preferences through historical post sequences and historical tag sequences. Finally, MLP4STR integrates the three parts for recommendation, which is more in line with the needs of tag recommendation. Thus, MLP4STR achieves the best results.

\subsection{Ablation Experiment}
In order to further study the influence of key modules in MLP4STR, this paper conducts ablation experiments to study the influence of sequence modeling module (MLP4STR), historical tag information (tag) and historical post information (doc) on the final performance, and the comparative evaluation results of F1@1, F1@3 and F1@5 on two of the datasets are shown in Figure 4.

After removing the MLP4STR module for sequence modeling and directly changing to the pooling layer, the F1@5 of the model decreases by 2.4\% and 1.4\%, respectively, which proves the effectiveness of the sequence modeling module for historical sequences. When only using user historical tag information or historical post information for modeling, each evaluation index decreases, which illustrates the role of using both historical tag information and historical post information for interaction modeling. Through the interactive learning of historical tag information and historical post information, user preferences are extracted, and the user's tag recommendation is better realized.

% 	图4  两个数据集上的消融实验F1@K对比结果

\begin{figure}[htbp]
        \centering
        \subfigure[Ablation experiment on cooking]
        {
            \begin{minipage}[b]{.45\textwidth}
                \centering
                \includegraphics[width=\textwidth]{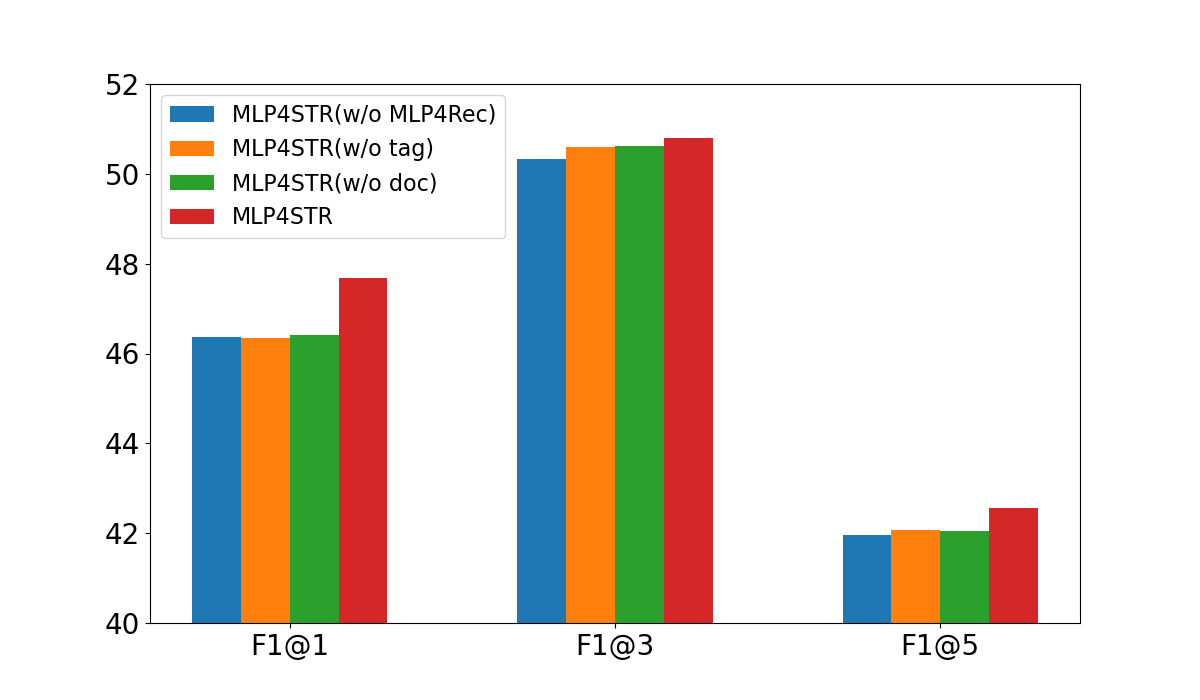}
            \end{minipage}
        }
        \subfigure[Ablation experiment on Android]
        {
            \begin{minipage}[b]{.45\textwidth}
                \centering
                \includegraphics[width=\textwidth]{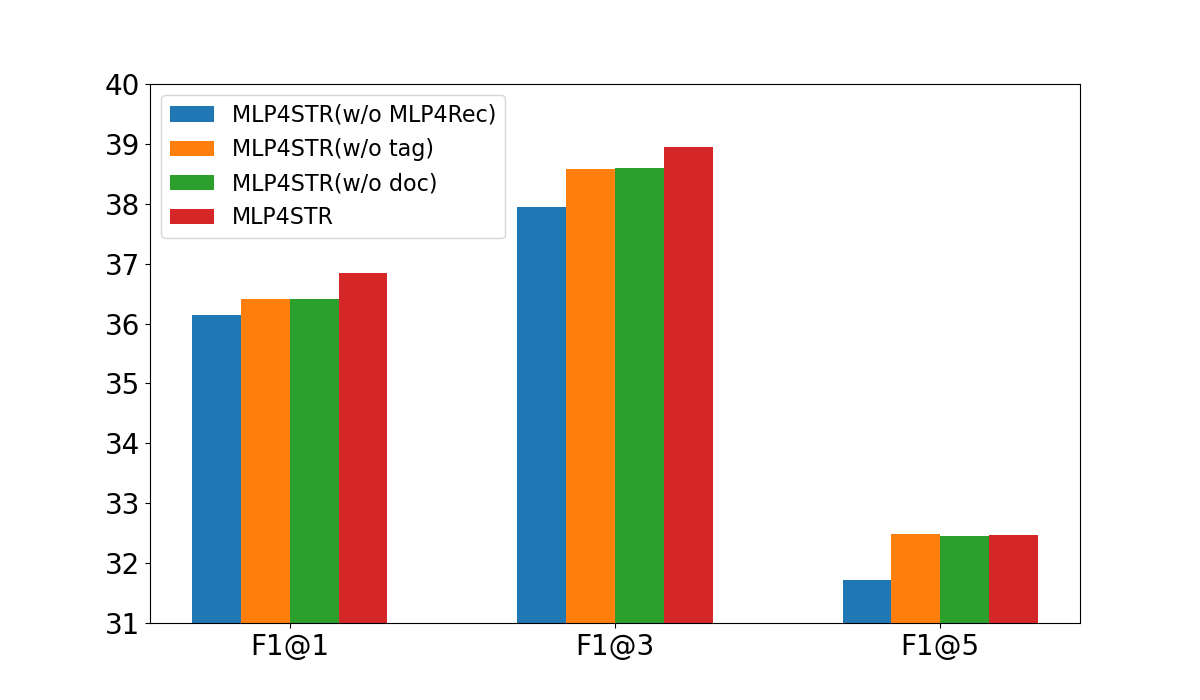}
            \end{minipage}
        }
 	\caption{F1@K comparison results of ablation experiments on two datasets}
 	\label{ablation}
    \end{figure}

\subsection{Parameter sensitivity analysis}
Since our model introduces the user's history information, in this section we analyze the effect of the length of the intercepted sequence of user's history information on the effectiveness of tag recommendation. The experimental results on both Cooking and Android datasets are shown in Fig.~\ref{ablation}
% 图5  两个数据集上的F1@5的变化
\begin{figure}[htbp]
        \centering
        \subfigure[Influence of hyper-parameters on Cooking]
        {
            \begin{minipage}[b]{.45\textwidth}
                \centering
                \includegraphics[width=\textwidth]{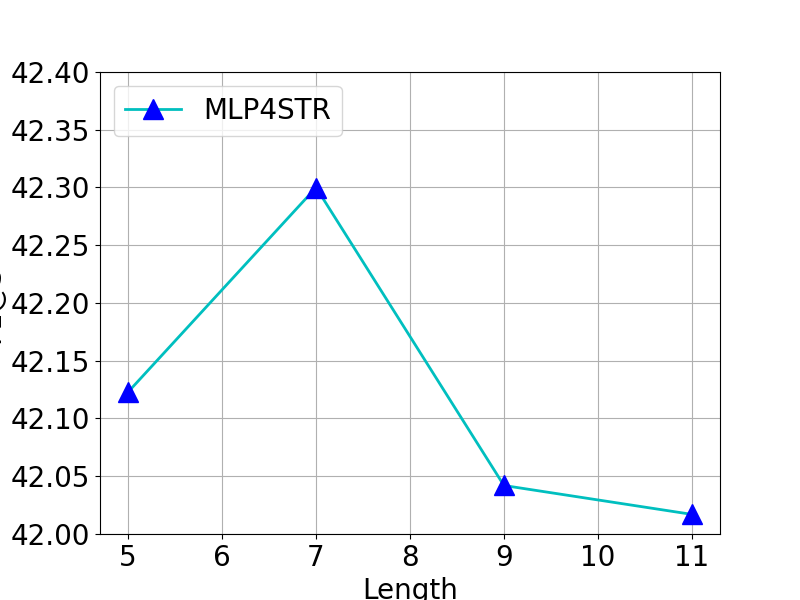}
            \end{minipage}
        }
        \subfigure[Influence of hyper-parameters on Android]
        {
            \begin{minipage}[b]{.45\textwidth}
                \centering
                \includegraphics[width=\textwidth]{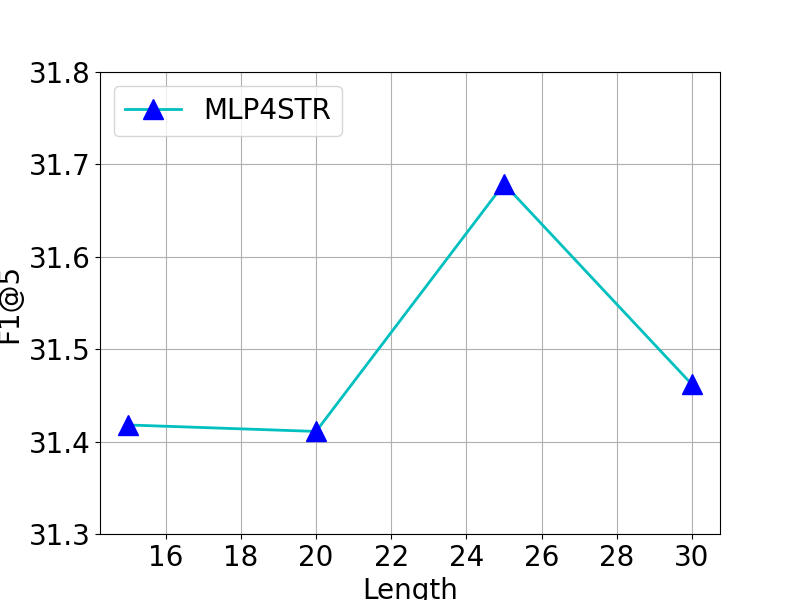}
            \end{minipage}
        }
 	\caption{Changes in F1@5 on the two datasets}
 	\label{ablation}
    \end{figure}

The average user history sequence length in the Android dataset is 26, and the results of F1@5 on this dataset peak at a history sequence length of 25. The average user history sequence length in the Cooking dataset is 8.2, and the results of F1@5 on this datamachine show an increasing and then decreasing trend around the average history sequence length.

\subsection{Case Study}

% 图6  标签推荐结果示例
\begin{table}
\vspace{-1.5em}
\begin{center}
	\caption{Example of tag recommendation results}
	\begin{tabular}{|p{10cm}|}
	\hline
	\textbf{Post1:} Does \textcolor{red}{Huawei-u8160} support USB host? I have a \textcolor{red}{Huawei-u8160} (A. K. AVodafone 858) which is running Cyanogenmod 7.2. I was wondering if it supports USB host so I can connect a keyboard or a Flash drive.\\
    \textbf{Tag:} cyanogenmod; usb-host-mode; \textcolor{red}{huawei-u8160};\\
    \textbf{Time:} 2012-06-21\\
	\hline
	\textbf{Post2:} where can I find fuse.ko file for my \textcolor{red}{Huawei-u8160} ? I have a \textcolor{red}{Huawei-u8160} running Cyanogenmod 7.2. I have been looking for the fuse.ko module for my phone, but whenever I find anything, terminal emulator tells me that it is incompatible! I want it to enable NTFS file system support using this tutorial So I was wondering where I can find the appropriate fuse.ko module.\\
    \textbf{Tag:} cyanogenmod; \textcolor{red}{huawei-u8160}; ntfs;\\
    \textbf{Time: }2012-09-16\\

	\hline
	\textbf{MLP4STR Result:} cyanogenmod; \textcolor{red}{huawei-u8160}; root-access;\\
	\hline
    \textbf{PTM4Tag Result:} cyanogenmod; root-access; partitions;\\
    \hline
	\end{tabular}
	\label{case}
	\end{center}
\vspace{-1.5em}
\end{table}

As shown in Table.~\ref{case}, Post2 is a post in the test set, Post1 is a post published by this user in history, and the graph contains the time information of the post published. It can be found that the method in this paper successfully recommends two tags cyanogenmod and huawei-u8160 for Post2, but only one tag is recommended by PTM4Tag. Although the content of the current post contains relevant content for the tag huawei-u8160, PTM4Tag is not able to recommend the tag.MLP4STR, after incorporating the sequence information, successfully extracts the user preferences presented in Post1 and uses them to guide the tag recommendation process for the current post.

\section{Conclusion}
In this paper, we propose the sequential tag recommendation model to deal with the tag recommendation problem, which uses a pure MLP structure based on cross-feature dimensions to interactively learn users' historical post sequences and historical tag sequences, and the learned user preferences are assisted for tag recommendation. In this paper, we propose for the first time the sequential tag recommendation task to obtain dynamic user preferences by sequentially modeling the user's historical information. Experimental results obtained on four real-world datasets show that the method proposed in this paper is able to effectively learn information about users' historical behavioral sequences in sequential tag recommendation.


\end{document}